\documentclass[12pt,fleqn]{article}
\usepackage{gc,amsfonts,amssymb,cite}

\def\mn{_{\mu\nu}}

\def\R{{\mathbb R}}
\def\S{{\mathbb S}}
\def\Z{{\mathbb Z}}

\def\bw{brane world}
\def\NL{Newton's law}

\def\Jl#1#2{#1 {\bf #2},\ }

\begin{document}
\prepno{gr-qc/0601114}{}

\section*{\Large\bf Brane world corrections to Newton's law}

\subsection*{K.A. Bronnikov$^{a,b}$, S.A. Kononogov$^{a}$
	and V.N. Melnikov$^{a,b}$}

\begin{description}\itemsep -1pt
\item[$^a$]
	{\small\it Center for Gravitation and Fundamental Metrology,
 	 VNIIMS\\ 46 Ozyornaya St., Moscow 119361, Russia}
\item[$^b$]
	{\small\it Institute of Gravitation and Cosmology,
 	 Peoples' Friendship University of Russia\\
 	 6 Miklukho-Maklaya St., Moscow 117198, Russia}
\end{description}

\subsection*{Abstract}

{\small
We discuss possible variations of the effective gravitational constant with
length scale, predicted by most of alternative theories of gravity and
unified models of physical interactions. After giving a brief general
exposition, we review in more detail the predicted corrections to Newton's
law of gravity in diverse brane world models. We consider various
configurations in 5 dimensions (flat, de Sitter and AdS branes in Einstein
and Einstein-Gauss-Bonnet theories, with and without induced gravity and
possible incomplete graviton localization), 5D multi-brane systems and some
models in higher dimensions.  A common feature of all models considered is
the existence of corrections to Newton's law at small radii comparable with
the bulk characteristic length: at such radii, gravity on the brane becomes
effectively multidimensional. Many models contain superlight perturbation
modes, which modify gravity at large scale and may be important for
astrophysics and cosmology. }

\section{Introduction \label{s0}}

In this paper, we are going to discuss one of the most important predictions
of a large and actively discussed class of modern theories of gravity,
related to different \bw\ scenarios, namely, possible violations of \NL\ of
gravity for macroscopic bodies. It seems useful to begin with giving a brief
but more general exposition of the to-date experimental and theoretical
situation with possible \NL\ violations.

Most of the existing generalized theories of gravity as well as
unified theories of basic physical interactions predict time and
also spatial variations of the gravitational constant
\cite{Mel2,Mel,02,St}which may manifest themselves as new
interactions in addition to \NL
\beq \label{m_1}
    	F = G m_1  m_2 / r^2.
\eeq
Such interactions can violate the equivalence principle (EP) if they depend
on the bodies' composition, modify the inverse square law and are often
described in terms of new particles --- new interaction carriers
\cite{m1,m2,m3,m4,m5}).

The laboratory experimental data and observations of planet and satellite
motion in the Solar system exclude, with great accuracy, any deflections
from \NL\ and, accordingly, the existence of new interaction carriers in
almost all length ranges, except the range much shorter than a millimeter
and the one from meters to hundreds of meters. This possible deflection may
be described by an additional Yukawa-type contribution to the
gravitational potential,
\beq                                                         \label{m2}
    	V = -(G m_1 m_2 / r) [1 +  \alpha \exp (- r/\lambda)],
\eeq
characterized by the interaction strength $\alpha$ and the scale $\lambda$
related to the mass of the interaction carrier. Among laboratory
experiments, there is only one result on a possible inverse square law
violation \cite{Italian} in the range of 20 to 500 m with the interaction
strength between 0.13 to 0.25. It was obtained by using a set of gravimeters
and an energy storage station, where water was pumped to a lake during the
night and then was released at daytime giving additional electric power.
Though, an independent verification of this result using other schemes is
absent.

More serious evidence on a possible violation of \NL\ has come to us from
space, namely, from data processing on the motion of the spacecrafts Pioneer
10 and 11, referring to length ranges of the order of or exceeding the size
of the Solar system. The discovered anomalous (additional) acceleration is
\cite{m7}
\[
       (8.60 \pm 1.34) \ten{-8}\ {\rm cm/s^2},
\]
it acts on the spacecrafts and is directed towards the Sun. This
acceleration is not explained by any known effects, bodies or influences
related to the design of the spacecrafts themselves (leakage etc.), as was
confirmed by independent calculations.

Many different approaches have been analyzed both in the framework of
standard theories and invoking new physics (Chongming Xu's talk at
ICGA-7, Taiwan, November 2005), but none of them now seems to be
sufficiently convincing and generally accepted. There are the following
approaches using standard physics:
\begin{description}
\item
    --- an unknown mass distribution in the Solar system (Kuiper's
    belt), interplanetary or interstellar dust, local effects due to the
    Universe expansion \cite{m6};
\item
    --- employing the Schwarzschild solution with an expanding boundary
    \cite{m7,m8} etc.
\end{description}

Among the approaches using new physics one can mention:
\begin{description}
\item
    --- a variable cosmological constant \cite{m9};
\item
    --- a variable gravitational constant \cite{m12};
\item
    --- a new PPN theory connecting local scales with the cosmological
    expansion \cite{m10};
\item
    --- the five-dimensional Kaluza-Klein (KK) theory with a
    time-variable fifth dimension and varying fundamental physical
    constants \cite{m11};
\item
    --- Moffat's \cite{m13} non-symmetric gravitational theory;
\item
    --- Milgrom's \cite{m14,m15} modified Newtonian dynamics (MOND);
\item
    --- special scalar-tensor theories of gravity \cite{m16};
\item
    --- approaches using some ideas of multidimensional theories
    \cite{m17};
\item
    --- modified general relativity with a generalized stress-energy
    tensor \cite{m18} etc.
\end{description}

This Pioneer anomaly has caused new proposals of space missions with more
precise experiments and a wide spectrum of research at the Solar system
length range and beyond:
\begin{description}
\item
  --- Cosmic Vision 2015-2025, suggested by the European Space Agency, and
\item
  --- Pioneer Anomaly Explorer, suggested by NASA \cite{m19}.
\end{description}

As to other theoretical schemes leading to \NL\ modifications,
from the general relativistic viewpoint, if we admit that the physical
constants may vary with time, it is quite natural to consider their spatial
variations, as was done in Ref.\,\cite{m2}.

As is well known, in Einstein's general relativity

\begin{description}
\item
      --- the gravitational interaction is carried by massless gravitons,
\item
      --- they are described by second-order differential equations, and
\item
      --- they interact with matter with an invariable strength related to
      the constant $G$.
\end{description}

If at least one of these features is violated, we shall in general arrive at
some distance-dependent \NL\ violation and consequently to a modification of
Einstein's theory. One can single out a number of classes of generalized
theories known in the scientific literature:

\begin{enumerate}
\item
   Theories with massive gravitons, e.g., bimetric theories.
\item
   Scalar-tensor theories with a variable effective gravitational constant
   \cite{St}.
\item
   Theories containing, in addition to the metric, space-time torsion and
   nonmetricity \cite{m2}.
\item
   Theories whose equations contain higher-order derivatives (e.g., due to
   quantum effects).
\item
   Theories with interaction carriers other than gravitons (the so-called
   partners): supergravity, superstrings, M-theory which promises to unify
   the latter but so far unfinished, etc.
\item
   Other nonlinear theories induced by any known interaction type, also
   creating effective non-zero masses \cite{m2}.
\item
   Various phenomenological theories which do not specify the cause of \NL\
   violation (the fifth force etc.)
\item
   Multidimensional theories: Kaluza-Klein and \bw\ approaches.
\end{enumerate}

All such theories lead to real or effective masses leading to additional
Yukawa or power-law corrections to \NL.

There are some model-dependent estimates of such forces. Thus, according to
Sherk \cite{Sherk79}, in supergravity where the graviton is accompanied by a
partner (graviphoton) of spin 1, there appears a Yukawa-type repulsion.
Moody and Wylcek \cite{MW84} introduced a pseudoscalar particle leading to
an additional attraction in the range between $2\ten{-4}$ and 20 cm with a
strength between 1 and $10^{-10}$. In Fayer's \cite{Fayer} supersymmetric
model a spin-1 graviphoton leads to an additional repulsion in a 10 km range
with a strength about $10^{-13}$. S. Weinberg's model \cite{Weinberg} with
a scalar field, aimed at generating a cosmological constant, predicted an
extra interaction at lengths smaller than 0.1 mm. Most of the
multidimensional theories \cite{Mel2,Mel,02}, including those developed and
analyzed in Ref.\,\cite{m2}, also predict \NL\ violations and the values of
post-Newtonian parameters other than in general relativity.

Most of the \bw\ models, to be discussed below in more detail, also predict
deflections from \NL\ at lengths smaller than 1 mm. At present, very active
experimental and theoretical studies are being performed in this range, see,
e.g., \cite{m5}. No \NL\ violations have been detected in the submillimeter
range by now \cite{Long}. It is expected that a few coming years will be
marked with an improvement of such estimates in the range from a nanometer
to a centimeter by several orders of magnitude, and hopefully such
experiments will be able to probe gravity in extra dimensions.

\section{The \bw\ concept \label{s1}}

The \bw\ concept, broadly discussed in the recent years, treats our
Universe as a distinguished three-dimensional (or four-dimensional if time
is included) surface or layer, called the brane, in a multidimensional
space-time where the extra dimensions are large or infinite. The Standard
Model fields are supposed to be concentrated on the brane while gravity
(and, as a rule, only gravity) propagates in the ambient space called the
bulk. The history of such models traces back to the early 80s, beginning
with Akama's \cite{akama}, Rubakov and Shaposhnikov's \cite{rubsh} and some
other works. The recent outburst of interest in such models is mostly
related to the progress in string and M-theory, in particular, with
Ho\v{r}ava and Witten's well-known 11D model \cite{h-witt}, in which one
of the extra dimensions is of much larger size than the others. This
approach has suggested, in particular, a natural mechanism for solving the
hierarchy problem in particle physics (Randall and Sundrum's model
\cite{RS1} and others), while retaining the Newtonian behaviour of weak
gravitational fields on the brane in agreement with modern experiments.
The \bw\ models have led to an appreciable progress in both particle
physics and cosmology.

By now, there are thousands of publications in this area, with a great
number of particular models and scenarios suggested, see, e.g., the reviews
\cite{rev-rub}--\cite{rev-coley}. The models differ in the following
essential properties:

\begin{itemize}
\item
    Total dimension $D$ (most of the studies consider $D=5$).
\item
    Models with single or multiple branes.
\item
    Flat or curved branes.
\item
    Flat or curved bulk.
\item
    Compact or non-compact bulk.
\item
    Possible fields other than gravity
    (spin 0, 1/2, 1) in the bulk.
\item
    Thin or thick branes. (An infinitely thin brane may only be
    considered as an approximation since any underlying theory contains
    a fundamental length, such that at smaller scales a classical
    description is meaningless.)
\item
    Various symmetries of the brane(s) and the bulk, etc.
\end{itemize}

Even this rough classification shows how diverse should be the
observational predictions: each model gives, in principle, effects of its
own.

Here we will consider the predictions of various \bw\ models for deviations
from \NL\ of gravity for material bodies situated (as well as the observer)
on the brane. We will be restricted, by necessity, to a certain set of most
elaborated models.

Laboratory measurements \cite{Long} have not discovered any deviations from
\NL\ at distances larger than 0.1 mm, which seriously constrains the
parameters of any generalized theories of gravity.

On the other hand, the astronomical observations show that \NL\ holds with
high accuracy at least up to stellar cluster scales. At the galactic scale,
however, there exists the well-known dark matter (DM) problem, and
modification of \NL\ \cite{MOND} is one of the ways of its solution. Some
\bw\ models also predict changes in \NL\ at kiloparsec and higher scales.
Some of these changes are able to remove (or at least ease) the DM problem
in galaxies and galaxy clusters. This set of problems is discussed in
Refs.\,\cite{rev-rub}, \cite{n1}--\cite{n9}.

\section{Modified \NL\ in RS2 model \label{s2}}

Apparently, the simplest \bw\ model (at least, the one containing a minimum
of details) is Randall and Sundrum's second model (RS2) \cite{RS2}, in
which a single brane, endowed with the Minkowski metric and a positive
tension, is embedded in a 5D anti-de Sitter (AdS) space-time with a constant
negative curvature. According to \cite{RS2}, the gravitational perturbations
of this model (gravitons) contain a zero (i.e, zero-mass) mode, concentrated
on the brane and responsible for the validity of \NL\ on the brane, and the
so-called Kaluza-Klein (KK) modes corresponding to all possible finite mass
values.

The RS2 model realizes a solution to the 5D vacuum Einstein equations with
the cosmological constant $\Lambda_5 < 0$ and a source in the form of a
$\delta$-like matter distribution in a certain spatial direction $y$.
The total cation has the form
\bear                                                      \label{S_RS2}
      S \eql S_{\rm grav} + S_{\rm brane},
\nn
      S_{\rm grav} \eql \int d^4 x \int dy\,\sqrt{|g_5|}
                    (2m_5^3\,R_5 - \Lambda_5),
\nn
      S_{\rm brane} \eql \int d^4 x \sqrt{|g_4|}\, V_{\rm brane},
\ear
where $R_5$ is the 5D scalar curvature, $g_5$ and $g_4$ are the determinants
of the 5D metric $g_{MN}$ and the 4D metric $g_{\mu\nu}$ at the hypersurface
$y = 0$, respectively; $m_5$ is the 5D Planck mass, which, in general, does
not coincide with the conventional one; the quantity $V_{\rm brane}$
describes the brane tension. The solution possesses a mirror ($\Z_2$)
symmetry with respect to $y=0$ and is described by the metric
\beq                                                          \label{AdS}
      ds_5^2 = \e^{-2|y|/l} \eta\mn dx^{\mu} dx^{\nu} - dy^2,
\eeq
(where $\eta\mn$ is the Minkowski metric, $l$ is the curvature radius of the
AdS bulk) and exists under the conditions
\beq
    V_{\rm brane} = 24 m_5^3/l, \cm \Lambda_5 = -24 m_5^3/l^2.
\eeq
Thus a ``fine tuning'' is required, which connects the input constants
$m_5$, $V_{\rm brane}$ and $\Lambda_5$ and providing a zero 4D cosmological
constant $\Lambda_4$. The 4D Planck mass value $m_4= 1/\sqrt{G}$ (where $G$
is the Newton constant of gravity) is related to the 5D constants:
$m_4^2 = m_5^3 l$.

Beginning with Ref.\,\cite{RS2}, a number of authors have calculated the
Newtonian potential on the brane with the result (for sufficiently large $r$)
\beq
    V(r) \approx \frac{GM}{r} [1+ \Delta(r)],      \label{V_RS2}
\eeq
and in all cases a correction of the form $\Delta \sim 1/r^2$ was obtained,
but with different numerical coefficients. A calculation of the scalar part
of the graviton propagator gave $\Delta (r) = l^2/(2r^2)$ at radii $r\gg l$
\cite{R4,R5}, whereas a calculation of the $h_{00}$ component of the metric
perturbation led to $\Delta (r) = 2 l^2/(3r^2)$ \cite{R6,R7,R8,R9}.
The difference between these results in the form of the factor 4/3 was
ascribed to ``brane bending'' by the point source.

Ref.\,\cite{n8} explains the advent of the factor 4/3 by the effect of the
full tensor structure of the graviton propagator, without need for effects
like  ``brane bending''.  The following expressions for the correction
$\Delta$ have been obtained ($\mu = 1/l$):
\bearr
    \mu r \ll 1:                                        \label{D_RS2}
  \cm
    \Delta =
    \frac{4}{3\pi \mu r} - \frac{1}{3} - \frac{1}{2\pi} \mu r \ln \mu r
          +  0.089237810\, \mu r + {\cal O}(\mu^2 r^2),
\nnn
    \mu r \gg 1:
  \cm
    \Delta = \frac{2}{3\mu^2 r^2} - \frac{4\ln \mu r}{\mu^4 r^4} +
          \frac{16 - 12\ln 2}{3\mu^4 r^4} +
            {\cal O} \!\left[ \frac{(\ln \mu r)^2}{\mu^6 r^6} \right].
\ear
For intermediate values of $r$, the correction was obtained numerically
and unifies these two asymptotic expressions. An approximate analytic
expression, valid for all $r$, was obtained in Ref.\,\cite{n3}.

Let us note that at small $r$ the main term of the expansion has the form
$\Delta \simeq 4/(3\pi\mu r)$. In this case, $\Delta \gg 1$, and the whole
potential behaves as $V \sim 1/r^2$: gravity becomes effectively
5-dimensional at radii smaller than the bulk curvature radius. The model
space in this approximation has almost the same properties as 5D Minkowski
space.

In this model, the curvature radius is $l \lesssim 0.1$ mm according to
the laboratory constraint \cite{Long}.

\section{Other brane models in 5D space \label{s3}}

\subsection{Branes in 5D Einstein-Gauss-Bonnet theory \label{s3.1}}

One of the generalizations of the RS2 models is a similar model in which
5D gravity is described, instead of general relativity, in the framework
of the 5D Einstein-Gauss-Bonnet theory (see \cite{n3} and references
therein) with the Lagrangian
\beq
     L'_{\rm grav} = R_5 - 2\Lambda_5                        \label{L_GB}
             + \alpha (R^2_5 - 4R^{LM} R_{LM} + R^{LMNP}R_{LMNP}),
\eeq
where $\alpha$ is a constant of dimension $l^2$, $R_5$, $R_{LM}$ and
$R_{LMNP}$ are the 5D scalar curvature, Ricci tensor and Riemann
tensor, respectively. At $\alpha=0$ this Lagrangian reduces to that of
general relativity (in notations slightly different from (\ref{S_RS2})).
The 5D metric in this model also has the form (\ref{AdS}), i.e., describes
an AdS space-time with the curvature radius $l$, but the perturbations,
which determine the form of corrections to the \NL, behave, in general,
differently.

The analytical expression obtained by Deruelle and Sasaki \cite{n3} for all
values of $r$ and $\alpha$, are very cumbersome and will not be presented
here, while their main conclusion is as follows: for values of the parameter
${\tilde\alpha} = 4\alpha/l^2$ close to unity, the gravitational potential
of an attracting centre reproduces the Newtonian potential much more
precisely than it is the case for an ``Einstein brane'' corresponding to
$\alpha = 0$ (in other words, the RS2 model). Namely, at small radii the
potential retains the form $\sim 1/r$, and consequently the modern
experiments do not constrain the curvature radius $l$ to values
smaller than 1 mm: the authors even assert that even values of $l$ of the
order of 1--100 km are compatible with the experiment. At distances $r \gg
l$, the $1/r$ dependence is also preserved, but with another value of the
gravitational constant. Since, according to the astrophysical data, the
constant $G$ at astronomical distances should not differ from the one
measured in laboratory by more than 10 per cent, this theory is subject to
the constraint $0.85 < {\tilde \alpha} < 1.15$ if $l$ is of the order
of 1--100 km.

\subsection{Inclusion of induced 4D gravity \label{s3.2}}

Some authors \cite{n5,n10,n11} consider RS2-like models, adding into the 4D
Lagrangian $L_{\rm brane}$ a term proportional to the 4D scalar curvature
$R_4$. It is explained as a contribution from one-loop quantum effects of
the matter fields existing on the brane. Outside the brane, gravity is
described by the 5D Einstein Lagrangian.

In the Dvali-Gabadadze-Porrati (DGP) model \cite{n11}, belonging to this
class and assuming a flat metric in the bulk, there appears a modification
of \NL\ at large distances, $r \gg r_c = m_4^2/m_5^3$, which is of potential
interest for explaining the observed acceleration of the Universe expansion.
To obtain such an effect, the radius $r_c$ should be comparable with the
Hubble radius $\sim 10^{25}$ m, and then the 5D Planck mass $m_5$ is
comparable with the proton mass. However, as shown by Rubakov \cite{n5}, at
such values of the parameters, the gravitational interaction turns out to be
anomalously strong at distances of the order of meters. This effect is caused
by the behaviour of the scalar (in the 4D sense) part of perturbations of
the 5D Minkowski metric. So the model \cite{n11} loses its appeal as a
possible explanation of the cosmological acceleration.

More general models with induced 4D gravity and an AdS bulk have been
analyzed in Ref.\,\cite{n10,sht05}. The action of the form \cite{sht05}
\bearr
	S = m_5^3 \biggl[ \int_{\rm bulk}                      \label{S_ind}
            (R_5 - 2 \Lambda_5) - 2 \int_{\rm brane} K\biggr]
\nnn \cm
     +\int_{\rm brane} (m_4^2 R - 2 \sigma) + \int_{\rm brane}L(h_{ab},\phi)
\ear
contains the constants $m_5$ and $m_4$ (the five-dimensional and
four-dimensional Planck masses, respectively), the bulk cosmological
constant $\Lambda$ and the brane tension $\sigma$. The quantity $K$ is the
trace of the symmetric tensor of extrinsic curvature of the brane, while
$L(h_{ab}, \phi$) denotes the Lagrangian density of the four-dimensional
matter fields $\phi$ confined to the brane and interacting only with the
its induced metric $h_{ab}$.

The corrections to \NL\ are of the same nature depend on the interplay of
scales related to the constants involved in (\ref{S_ind}) and generalize
those of the RS2 and DGP models. It is shown, in particular, that, for some
values of the constants, the $\sim 1/r$ dependence of the potential is
preserved at small radii \cite{n10}, similarly to the results \cite{n3} for
branes in Einstein-Gauss-Bonnet gravity.

\subsection{Curved branes in 5D space-time \label{s3.3}}

So far we have been describing \NL\ modifications on Minkowski branes, i.e.,
4D gravity was considered as a small perturbation in a flat background.
This approach is justified by that the brane curvature (in other words, the
observed space-time curvature) is small as compared to the curvature of
multidimensional space which leads to the formation and existence of the
brane itself. Many authors nevertheless consider curved branes, most
frequently de Sitter (dS) branes. This maximally symmetric space-time is
attractive observationally since it describes a Universe in accelerated
expansion and are of interest from a theoretical viewpoint, in particular,
due to the conjectural dS/CFT correspondence \cite{Hull,Stro}.

Refs.\,\cite{Od02, R5} (see also references therein) calculate the
gravitational potential on a dS brane by the perturbation method, assuming
three forms of bulk space-times: dS, Minkowski and AdS. In the first case
\cite{Od02}, for distances $r \gg l$ (where, as before, $l$ is the curvature
radius of the AdS bulk), the result is the same as in the RS2 model:
$\Delta = 2 l^2/(3 r^2) + o(l^2/r^2)$ --- cf. \eqs (\ref{V_RS2}) and
(\ref{D_RS2}). At small $r$ the potential also behaves
``five-dimensionally'', i.e., $\sim 1/r^2$. It is of interest that the brane
curvature, whose radius is assumed to be much larger than $l$, does not
affect the effective Newtonian potential.

A different behaviour of the gravitational perturbations is found for dS
branes in dS bulk \cite{Od02}.  First, at $r\gg l'$ (where $l'$ is the
curvature radius of the 5D de Sitter space) the correction is negative:
$\Delta = -2 {l'}^2/(3 r^2) + o(l^2/r^2)$.  Second, tachyonic perturbation
modes are revealed, which are probably related to instabilities of the
initial solution to the gravitational field equations.

Ref.\,\cite{R5} describes another result for 5D Minkowski and dS spaces: it
is claimed that it is only possible to calculate the corrections in the near
zone, i.e., for small radii, and the corrections exceed the basic Newtonian
potential related to the zero graviton mode. It is concluded that only AdS
space-time is suitable for describing the bulk. A similar conclusion is made
in Ref.\,\cite{Tamv} in an analysis of gravitational perturbations
outside a dS brane.

It should be noted that the cited papers have only discussed thin branes for
which the corresponding solution of 5D gravity equations are easily
found. Refs.\,\cite{thk2, thk3} have shown that a Minkowski brane
treated as a thick domain wall, appearing in 5D space due to the violated
$\Z_2$ symmetry of a scalar field, can only exist in an asymptotically AdS
bulk; in this case, the RS2 model with a thin brane is obtained in a
universal way (independently of the shape of the symmetry breaking
potential) as a properly defined limit. These results correlate with that of
\cite{R5}: it is, e.g., natural to suppose that a weakly curved dS brane
with the curvature radius $R$ cannot exist in a 5D dS bulk with the
curvature radius $l \ll R$. This apparently means that those solutions of
the 5D gravity equations for thin branes which are, according to
Refs.\,\cite{Od02, R5, Tamv}, unstable or unrealistic, simply cannot be
obtained as well-defined limits from thick brane solutions.

\subsection{Brane world in dilaton gravity \label{s3.4}}

Among the multidimensional gravity theories assuming other fields in the
bulk in addition to the metric, the simplest one is apparently the theory
with a single scalar field which is sometimes called dilaton gravity. The
corresponding 5D Lagrangian is
\beq                                                          
    L_{\rm DG} = R_5 + \Half g^{AB} \d_A\phi \d_B\phi - V(\phi),
\eeq
where $V(\phi)$ is the dilaton potential. Ref.\,\cite{od00} has shown that
there are \bw\ solutions with the bulk metric
\beq                                                          \label{ds_dil}
      ds_5^2 = \e^{2A(y)} \gamma\mn dx^{\mu} dx^{\nu} - dy^2,
\eeq
where the 4-metric $\gamma\mn$ may be, in particular, Minkowski, dS or AdS
while the functions $A(y)$ and $\phi(y)$ are determined from the 5D dilaton
gravity equations. The inclusion of a scalar is motivated by that the
graviton is accompanied by scalar partners in all models of string and
M-theory and by considerations related to the AdS/CFT correspondence.
As in the RS2 model, the brane is situated at a fixed hypersurface
$y = \const$, at which the function $A(y)$ suffers a fracture. It has been
proved that gravity can be localized on the brane in the limit of a small 4D
cosmological constant, and the corrections to \NL\ due to KK models have been
calculated. The modified Newtonian potential on the brane has the form
(\ref{V_RS2}) for sufficiently large radii \cite{od00}, but, unlike the
previously discussed models, the correction has the form
$\Delta \sim 1/r^3$, with a coefficient depending on the choice of the
specific solution. The additional potential is thus proportional to
$r^{-4}$ instead of $r^{-3}$ in other model, making gravity stronger at
small radii.

For a similar class of models with an exponential potential $V(\phi)$,
Ref.\,\cite{V01} has studied all possible tensor and scalar perturbations.
Apart from the corrections to the Newtonian potential (which, in a certain
special case, coincides with the RS2 result), an effective scalar-tensor
interaction on the brane has been discovered. This interaction contains both
short-range and long-range components and has been estimated as a
potentially dangerous one for the \bw\ scenario.

\subsection{Incomplete localization of gravitons \label{s3.5}}

Some authors consider the possibility of obtaining phenomenologically
acceptable models with infinite extra dimensions under the assumption that
the 5D gravitons form, instead of a usual zero mode, a metastable bound
state with a small but finite probability of leaving the brane.

One such model was studied in the papers by Gregory, Rubakov and Sibiryakov
\cite{GRS1,GRS2} (see also references therein). In 5D space-time,
one assumes the existence of three parallel branes: the central one,
with the positive tension $\sigma$, and two identical lateral ones with
the negative tension $-\sigma/2$, at equal spacings from the central brane.
The whole system possesses $\Z_2$ symmetry. The usual matter is located
on the central brane. The cosmological constant $\Lambda_5$ is negative
between the branes and is zero outside them. At proper values of
$\Lambda_5$, there are solutions to the 5D Einstein equations with the AdS
metric between the branes and the flat metric outside. A linear perturbation
study reveal the following: 1) There is an intermediate range of distances
where the 4D Einstein equations hold (and consequently the usual Newton
law). 2) Phenomenologically acceptable bounds of this range may be obtained
by a proper choice of the model parameters and, as is claimed, ``without
strong fine tuning''. 3) At both small and large distances, there are
contributions to the gravitational potential $\sim 1/r^2$. For small
distances, as in other models, it means a transition to an effectively 5D
nature of gravity. 4) At very large (cosmological) distances, there appears
a contribution of the form $G_4/(3r)$, where $G_4$ is Newton's constant from
the intermediate range. Thus at large $r$ the gravitational field turns out
to be repulsive and represents ``scalar antigravity''.

Ref.\,\cite{n9} has studied tensor perturbations of sufficiently general 5D
models of flat branes, with the metric
\beq                                                      \label{ds_n9}
    ds^2 = \e^{-A(z)} (\eta_{\alpha\beta}dx^\alpha dx^\beta -dz^2),
\eeq
where $A(z)$ is an even function, non-decreasing at $z > 0$. The graviton
zero mode turns out to be quasi-localized if $A(z)$ grows sufficiently
slowly at large $z$, in particular, if it tends to a constant, which happens
if the metric (\ref{ds_n9}) has a flat asymptotic. It has been shown that
 the conclusion \cite{GRS2} on the existence of an intermediate range of
radii with Newtonian gravity is general for all such models. As to the
behaviour of the effective gravitational potential at large distances,
its asymptotic has the form $1/r^{1+\alpha},\ 0 < \alpha \leq 1$, where
the constant $\alpha$ depends on the model parameters and is equal to 1
for an asymptotically flat 5-geometry. However, the contribution of the
scalar sector of the metric perturbations was not analyzed; meanwhile, it
was this sector that led in Ref.\,\cite{GRS2} to an antigravitational
potential at large radii. The authors of \cite{n9} stress that this sector
does not follow such universal laws as the tensor perturbations and
requires consideration in a more specific problem setting.

\subsection{5D models with multiple branes \label{s3.6}}

In this class of models, the most well-known is Randall and Sundrum's first
model (RS1) \cite{RS1}, containing two branes and providing a solution to
the interaction hierarchy problem due to the exponential factor in the AdS
metric (\ref{AdS}) in the bulk. The extra dimension is compact and has the
form of a ring, and the branes are located at its antipodal points. In the
RS1 ``scenario'', the Standard model fields of particle physics are assumed
to be localized on the ``Planckian'' brane with negative tension while the
other (TeV) brane, possessing a positive tension, may contain ``shadow''
matter which is invisible from the observable Planckian brane. A
characteristic feature of such models is the existence of perturbations in a
mode related to a variable distance between the branes, called the radion.
The latter behaves in an effective 4D theory as a scalar field which does
not directly interact with matter but makes an appreciable contribution to
the predicted law of gravity.

The effect of the radion and a possible shadow matter on the observable
gravitational field has been analyzed, in particular, in
Refs.\,\cite{R6,smol02,arn04,sht03} (including RS1 and other two-brane
models). It has been shown \cite{R6} that, subject to the radion, a weak
field on each brane is described by the linearized Brans-Dicke theory of
gravity with different parameters $\omega$: $\omega > 0$ on the
positive-tension brane, and $\omega > 3000$ if the brane separation exceeds
the AdS curvature radius by a factor of four. \NL\ holds on each brane but
with different effective gravitational constants and with complicated
corrections at both small and large radii. The gravity of shadow matter from
the other brane is also felt, and, in some variants of the theory, even
stronger than that on the observed brane. Similar results have been obtained
in Ref.\,\cite{smol02}, but with other numerical values of some
coefficients. It has been shown \cite{arn04} that gravity of point particles
on the TeV brane is much stronger than on the Planckian brane. Thus the
results of different authors slightly vary, and, probably, some further
analysis will follow in order to remove the discrepancies. Of particular
interest is, however, the possible existence of shadow matter and its
gravitational interaction with the observed matter. Though, some arguments
have been put forward for the non-existence of any matter on the second
brane \cite{smol02,smol04}. On the other hand, it has been shown
\cite{smol04, smol05} that in some RS1 type models (including 4D gravity of
each brane, i.e., with induced 4D gravity, as in \sect \ref{s3.2}), the
radion may be absent as a result of a certain additional symmetry between
the branes.

Rather general two-brane models with AdS bulk have been studied in
Ref.\,\cite{shta03}: the action is written in a form like (\ref{S_ind}) but
the brane integrals encompass two branes with different values of $m_4$ and
$\sigma$. Negative values of $m_5$ and the brane tensions are allowed.
The gravitational fields are calculated for sources placed on any of the two
branes. As in other models, there is a vast range of sufficiently large
radii where \NL\ holds (with the effective gravitational constant depending
on the model parameters), and there appear corrections $\sim r^{-2}$ and
$\sim \ln r$ at small radii; their nature and ranges again depend on the
interplay of the parameters.

As is noted in \cite{shta03}, exploring both signs of $m_5$, related to the
bulk gravitational constant, and negative brane tensions may be of interest
in cosmological applications, such as models with disappearing dark energy.

Configurations with a larger number of branes are described in detail in the
PhD thesis by Mouslopoulos \cite{n1}, written on the basis of studies
conducted in the Oxford group, see, e.g., \cite{ox1}--\cite{ox4}.

A new phenomenon of interest that appears in multiple-brane systems is
the so-called multi-localization. It occurs if the potential $V(z)$ in the
corresponding Schr\"odinger equation (for any physical fields and, in
particular, for gravitational perturbations of the background model)
contains at least two wells. In a quantum-mechanical description, a brane
with a positive tension $\sigma > 0$ creates a $\delta$-function shaped
potential well while a brane with $\sigma < 0$ creates a similar potential
barrier. The possible tunnelling in such models leads to nontrivial spectra
of KK states including superlight localized KK states. Tunnelling removes
the degeneracy of zero modes and creates an exponentially small splitting in
the mass spectrum. Other levels, which do not correspond to bound states,
form a usual KK spectrum.

Multi-localization is of particular interest because it creates, in a
system with a single mass scale existing from the outset, another,
exponentially smaller mass scale.

In the simplest case, we obtain a picture in which the observed
gravitational interaction is a summed effect of a massless graviton and a
massive super-light KK state. A large energy gap between the first KK state
and the remaining tower leads to the existence of Newtonian gravity at
intermediate scales. A radical prediction of the model, apart from the
short-distance modification of gravity (owing to ``heavy'' KK modes), is its
modification at superlarge distances due to a superlight KK mode with
nonzero mass \cite{ox4}.

In more complex multi-brane systems there can be multiple superlight modes
making the predictions more complex and diverse. However, as noted by the
authors themselves \cite{n1}, this class of models either contains a
``ghost'' radion scalar field (with a wrong sign of energy) or it is
necessary to introduce a negative cosmological constant $\Lambda_4$
on the observed brane that leads to the AdS geometry, whereas the
cosmological observations are only compatible with $\Lambda_4 > 0$. The
authors hoped to get rid of these difficulties by considering
higher-dimensional models \cite{ox2}.

\section{Models with multiple extra dimensions \label{s3.7}}

Models with more than five dimensions are also various and also predict
essential changes in \NL\ at small distances. There is a common physical
reason for such a behaviour: the higher dimension manifests itself at scales
comparable with the bulk curvature radius.

In Roessl's PhD thesis \cite{ro-thes}, which completes a number of works
with the author's participation (see, e.g., \cite{ro1,ro2}),
multidimensional models with a single brane are classified as follows:

1) ``Strictly local'' branes, i.e., branes in which the stress-energy tensor
related to brane matter is either zero in the bulk or decays at least
exponentially.

2) Global topological defects (strings, monopoles) with sets of scalar
fields $\phi^a$ of sigma-model type with ``hedgehog'' configurations
($\phi^a = \phi(x) n^a$, $ n^a n^a =1$) and a spontaneously broken global
symmetry, e.g., $O(N)$ \cite{ro27,ro124,ro130}.

3) Models similar to the `t Hooft-Polyakov magnetic monopole, with a gauge
field in extra dimensions \cite{ro2}.

The cited papers considered $D=4+n$-dimensional metrics of the form
\beq                                                          \label{ds_D}
     ds_D^2 = A(l) \eta\mn dx^\mu dx^\nu + dl^2 + R^2(l) d\Omega^2_{n-1},
\eeq
where $\eta\mn$ is the 4D Minkowski metric while $d\Omega^2_{n-1}$ is
the linear element on a unit $(n-1)$=dimensional sphere. In the
extra-dimensional spherical coordinates, the brane is situated at the
centre, $l=0$. The following expression for the Planck mass was used:
\beq                                                          \label{M_P}
    M_{\rm P}^2 = {\cal A}_{n-1} M_D^{n+2}
            \int_{0}^{\infty} A(l)\, R^{n-1}(l)\, dl,
\eeq
where ${\cal A} = 2\pi^{n/2} /\Gamma(n/2)$ is the ($n-1$)-dimensional area
of a unit sphere. The finiteness of the expression (\ref{M_P}) served
as a criterion for localization of gravity.

According to \cite{ro-thes}, under the ``strictly local'' assumptions, no
acceptable multidimensional solutions have been found, and it is asserted
that this class does not contain any (at least simple) models with localized
gravity.

In models with global topological defects \cite{ro126,ro1} with a
Mexican-hat potential for the scalar fields, $V = \lambda(\phi^a\phi^a
- \eta^2)^2$ where $\lambda$ and $\eta$ are constants, the authors sought
models with the function $A(l)$, decaying far from the brane by the law
$\e^{-cl}$. Such a solution was found, such that the spherical radius
$R\to \const$ at large $l$, i.e., the extra dimensions form an
$n$-dimensional cylinder $\R_+ \times \S^{n-1}$. It is found that such a
scalar structure is only necessary for obtaining solution with a decreasing
function $A(l)$ for $n\geq 3$. It leads, just as the 5D models, to
corrections to \NL\ of the form $\sim 1/r^2$, which is physically explained
by the fact that only one extra dimension remains non-compact.

At other value of the parameters, one obtains singular solutions with
$R\to 0$ as $l \to \infty$. Assuming that the singularity may be smoothed
away by higher string corrections, a modified Newtonian potential was also
calculated for this model, with the result
\beq
    V(r) \approx \frac{GM}{r}              \label{V_RS2'}
        \left [1 + \frac{\Gamma(n+2)} {2\Gamma^2[(n+3)/2]}\,
              \frac{1}{(cr)^{n+1}}\right],
\eeq
where the correction rapidly decays at large $r$ and with growing number of
dimensions $n$.

The following remark is in order here. In the above model, the ``cylinder
end'', which is infinitely remote in the static reference frame, is
accessible for geodesics at finite proper time and represents a killing
horizon at which the function $A$ turns to zero. Consequently, the spatial
part of the 4D metric $A \eta\mn$ becomes degenerate, and an arbitrarily
small matter density reaching this region should turn to infinity at
such a horizon, leading to a space-time singularity \cite{bm05}. This actual
instability leads, in our view, to serious doubts that this and similar
models can be viable.

The third type of multidimensional models \cite{ro2} includes gauge
$p$-forms with the strength tensors $F_{A_1\ldots A_{p+1}}$ in the bulk,
where $p$, the rank of the form, may be different, up to $D-1$ (in the
latter case, however, they only create an addition to the $D$-dimensional
cosmological constant). The necessity of considering such models is
motivated by the inevitable emergence, in the case of global symmetry
violation, of a massless Nambu-Goldstone boson, leading to difficulties with
stability \cite{ro-thes}. Ref.\,\cite{ro2} describes a model with $p = n-2$
and a regular cylinder-like geometry in the extra dimensions, with a
constant radius $R$ as $l\to \infty$. It is a direct generalization of the
't Hooft-Polyakov monopole. Corrections to \NL\ are proportional to
$1/r^2$, as in similar models with global monopoles and for the same reason.

\section{Concluding remarks}

This review briefly describes the corrections to \NL\ for only some most
well-known \bw\ models, and the authors bring their apologies to the
colleagues whose work, being maybe not less interesting and important, is
not mentioned here.

With all the diversity of models, the nature of the corrections is more or
less common: at small distances between the gravitating bodies, close to
the bulk curvature radius, the world's multidimensional geometry comes to
the scene. For the number of dimensions higher than five, one could expect
a universal dependence of the correction on the total dimension $D$. The
analysis has shown, however, that it is the number of non-compact extra
dimensions that is essential rather than the total dimensionality.

Of special interest are apparently the models with superlight metric
perturbation modes which predict a modification of the gravitational
interaction for matter on the brane at astronomical scales. Such corrections
to Newton's gravity may be promising for resolving the problem of dark
matter (DM) in galaxies and galaxy clusters and even the cosmological dark
energy problem.

This is, however, only one of many approaches to these problems. Thus, apart
from the well-known attempts to attribute the DM to dark celestial bodies
and weakly interacting massive particles (WIMPs) or to the modified laws of
gravity in four dimensions \cite{MOND,kirv}, the DM effect may be explained
in a different way in the \bw\ framework. Thus, it has been shown \cite{pal}
that the astrophysical observations which are usually ascribed to DM can be
explained by extra terms in the effective Einstein equations for gravity on
a curved brane of RS2 type \cite{SMS}. It is also claimed \cite{pal} that
future gravitational lensing observations will be able to discriminate
between different explanations of the DM effects.

It should be noted that the enumerated models in which the coefficients by
$\eta\mn$ (frequently called warp factors) decay as one moves away from the
brane, so successfully used for the description of localized gravity,
are not free of problems with non-gravitational matter localization
\cite{thk3,bm05,gog,gog-sing}. It has been shown, in particular, that in
all $\Z_2$-symmetric 5D models with a single brane (i.e., all RS2 type
models), any scalar field mode has an infinite energy per unit volume on the
brane since the corresponding integral over the fifth dimension diverges
\cite{thk3}. Such problems emerge owing to the repulsive (from the brane)
nature of the bulk gravity: the warp factor is actually a multidimensional
generalization of the conventional gravitational potential, and particles
tend to rolling down to its minimum, the AdS horizon in the present case. In
our view, it makes sense to analyze, more thoroughly than it has been done
so far, the \bw\ models that are free of these problems, namely, those with
attractive gravity in the bulk. Some such models are described in
Refs.\,\cite{gog,gog-sing,bm05}.

\end{document}